\documentclass[journal]{IAENGtran}

   \usepackage[outdir=./]{epstopdf}
   \usepackage[pdftex]{graphicx}
   \DeclareGraphicsExtensions{.pdf}

\usepackage[cmex10]{amsmath}
\usepackage{amsthm}
\usepackage{amssymb}

\begin{document}
%
\title{Nondeterministic functional transducer inference algorithm}
%
%
%

\author{Aleksander Mendoza-Drosik}

\newtheorem{theorem}{Theorem}
\newtheorem{definition}{Definition}

\maketitle

\pagestyle{empty}
\thispagestyle{empty}

\begin{abstract}

The purpose of this paper is to present an algorithm for inferring nondeterministic functional transducers. Polynomial procedure is achieved by additionally  assuming that transducers are locally prefix-preserving. The algorithm is a generalisation of two other well known algorithms in the filed: RPNI and OSTIA. Functional transducers are all those nondeterministic transducers, whose regular relation is a function. Locally prefix-preserving transducers assume, that for any two paths starting in initial state, if input and output of first path is a prefix of the respective input and output of the other path, then the first one is a subpath of the other. Epsilon transitions as well as subsequential output can be erased for such machines, with the exception of output for empty string being lost. Learning partial functional transducers from negative examples is equivalent to learning total ones from positive-only data. 
\end{abstract}

\begin{IAENGkeywords}
functional transducers, ostia, rpni, nondeterminism, grammatical inference
\end{IAENGkeywords}

%
\IAENGpeerreviewmaketitle

\section{Introduction}
%
%
%
%
\IAENGPARstart{L}{earning}
of nondeterministic automata has always been a topic of great interest, although not many positive results were achieved. Most of research focused on weighted automata\cite{DROSTE} and probabilistic machines\cite{MOHRI}\cite{MOHRI3}. Algorithms like APTI\cite{HANSAN} allowed for learning transducers from distribution. Some attempts at generalising non-probabilistic machines were also made, such as the semi-deterministic transducers\cite{semideterministic}. More results\cite{activeLearningNondetFST} were obtained by using active learning and queries. Relatively few research has been done that would attempt to learn nondeterministic automata from text only. In the general case it can be proven that such a task is impossible. Only so far known positive results were for algorithms like OSTIA\cite{OSTIA}, RPNI\cite{RPNI} and its derivatives, but they assumed determinism. The algorithm in this paper presents a generalisation of the two previous algorithms that relaxes assumption of determinism. Here we only assume the transducer to be functional\cite{TRANSDUCERS}\cite{MendozaDrosik2020MultitapeAA} and locally prefix-preserving. 

\section{Assumptions}

The task is to learn functional nondeterministic transducers from informant. Their transition function is defined as $\delta: Q \times \Sigma \times Q \rightarrow \Gamma^*$ where $Q$ is set of states, $\Sigma$ is the input alphabet and $\Gamma$ is the output. All the formal relations that can be expressed by functional transducers are of the form $\Sigma^* \rightarrow \Gamma^*$. We do not allow subsequential transducers (only transitions have outputs but states don't) and neither allow for $\epsilon$-transitions. As a result the empty string $\epsilon$ can only be in relation with another empty string $(\epsilon,\epsilon)$. Hence pairs of the form $(\epsilon,\gamma)$ for any non-empty $\gamma$ will never appear in informant. It's worth pointing out that in the case of such functional transducers, the subsequential output, as well as, $\epsilon$-transitions can be erased and reduced to transducers without them\cite{MendozaDrosik2020MultitapeAA}.

We can assume that the transducers are total for all non-empty strings (that is, the rational relation $\Sigma^* \rightarrow \Gamma^*$ is total, except for $(\epsilon,\epsilon)$, which may or may not belong to the relation), because every partial nondeterministic functional transducer can be reduced to a total one. Such reduction is done by adding some new special symbol $\#$ to the alphabet $\Gamma$ and creating a new total transducer that returns $\#$ for all inputs that would otherwise be rejected by the partial transducer. More precisely, it can be achieved by taking the input projection (accepted subset of $\Sigma^*$ for which partial transducer returns output), turning it into DFA, negating it and then turning it back into transducer by making it return $\#$ for all accepted inputs (except $\epsilon$). Lastly we need to perform union of this new negated transducer with the original one. Therefore it's possible to encode counterexamples only by using informant consisting of pairs $\Sigma^* \rightarrow \Gamma^* \cup \{\#\}$. This proves that learning of partial transducers from negative examples is reducible to learning total transducers from only positive examples and vice versa. It's worth pointing out that such reduction would not be possible for deterministic transducers (due to preservation of prefixes\cite{MendozaDrosik2020MultitapeAA}, which we define below).

Informant is defined as infinite sequence of pairs $\Sigma^* \times \Gamma^*$. Because the transducers are total, eventually every string from $\Sigma^+$ will appear in the informant. Because transducers are functional, every such  $\Sigma^*$ string uniquely determines $\Gamma^*$ output that appears along with its input in the informant. (For this reason, functionality also implies that output of each transition can be uniquely determined by its source state, target state and input symbol.) In order words, once we see pair $(\sigma,\gamma)$, we can be sure that next time we encounter $(\sigma,\gamma')$, the outputs will be the same $\gamma=\gamma'$.  During learning, the algorithm only has access to some finite initial segment of the informant, but we can make it as large as necessary. The learning will converge to some correct hypothesis in the limit, as size of this segment approaches infinity.

Before showing the algorithm, let's first prove that learning in the limit is possible for functional nondeterministic transducers. This can be done, by observing that finding the minimal transducer consistent with any finite part of informant is computable. We can enumerate all transducers starting from the small ones and slowing moving onto the larger ones, until we eventually find one that returns expected outputs for all inputs. Suppose there is some other transducer with no more states than the target transducer $T$, which we're trying to learn. If both transducers determine the same regular relations, then it doesn't matter which one we infer. However, if they are different, then we will at some point find a pair in the informant that tells them apart (transducer being functional, is the key here) and the inference algorithm will make a mind change. Because there are only finitely many automata smaller or equal to the target transducer, there will be only finite number of mind changes before reaching the correct hypothesis. Hence learning will always converge to some equivalent minimal transducer.

Such algorithm, is simple but not very practical. A polynomial procedure can be achieved by making one further restriction. We need to assume that the automata are not only functional but also locally prefix-preserving. Preservation of prefixes is a property that for every two pairs $(\sigma_1,\gamma_1)$ and $(\sigma_2,\gamma_2)$ recognized by transducer, states that if $\sigma_1$ is prefix of $\sigma_2$ then $\gamma_1$ is a prefix of $\gamma_2$. Deterministic transducers preserve prefixes. Local preservation of prefixes is a relaxed version of this property. Let $p_1$ and $p_2$ be any two paths starting in initial state. Let $(\sigma_1,\gamma_1)$ be the input and output of $p_1$ obtained by concatenating all consecutive input and output labels from all transitions in $p_1$. Analogically we define $(\sigma_2,\gamma_2)$ for $p_2$. Locally  prefix-preserving transducer guarantees that for any $p_1$ and $p_2$, if $\sigma_1$ is a prefix of $\sigma_2$ and $\gamma_1$ is a prefix of $\gamma_2$, then the $p_1$ is a prefix-path of $p_2$. This property implies that transducer must be unambiguous, that is, for any accepted input, there is only one possible accepting path. 

Unambiguity  doesn't reduce the expressive power of automata, because every functional transducer can be converted to unambiguous one. The proof is simple and similar to powerset construction. Given some functional transducer $T$ with states $Q$,  build a new one $T'$ with set of states $Q' = Q \times 2^Q$. All transitions in original transducer $T$ are of the form $\delta(q_1,\sigma,q_2)=\gamma$ and $\delta$ is a partial function. We put a transition $\delta'(q_1',\sigma,q_2')=\gamma$ between $q_1' = (q_1,K_1)$  and $q_2' = (q_2,K_2)$,  whenever $\delta(q_1,\sigma,q_2)=\gamma$ and $\hat{\delta}(K_1,\sigma)=K_2$, where $\hat{\delta}$ is the image of $\delta$ defined as $\hat{\delta}(K_1,\sigma)=\{q_2\in Q : \delta(q_1,\sigma,q_2)\ne\emptyset \}$. The state $q_1'=(q_1,K_1)$ is accepting, whenever $q_1$ is accepting. At this point, the obtained powerset automaton is equivalent to the original one, but not unambiguous yet. The last step is to drop some of the transitions that are not necessary. If there are two transitions coming to the same $q_2'$ over the same symbol $\sigma$, they both must "carry" with them the same output (otherwise transducer wouldn't be functional). Hence one of them can be arbitrarily deleted. Analogically, if there are two states 
$q_1'=(q_1,K_1)$ and $q_2'=(q_2,K_2)$, such that $K_1=K_2$ and both $q_1$ and $q_2$ are accepting, then we don't need to make both $q_1'$ and $q_2'$ accepting. This finishes the conversion.

It's worth noting that due to unambiguity, every element $(\sigma,\gamma)$ in the informant uniquely determines exactly one accepting path in the target transducer. Lastly, even though unambiguity doesn't reduce the expressive power, the assumption of local preservation of prefixes does. It's not a significant limitation, because the class is still strictly larger than that of deterministic subsequential transducers and it also includes most of the relations recognizable by nondeterministic functional transducers.

\section{Initialization}

The inference algorithm needs to be initialized with maximal canonical prefix tree automaton, but due to nondeterminism, its construction is a little different from OSTIA or RPNI. Every state $q$ of the prefix tree corresponds to some state $\bar{q}$ of the original transducer $T$ that we are trying to learn (but the algorithm doesn't know $\bar{q}$). By  $\mathcal{L}(\bar{q})$ we will denote the relation defined by state $\bar{q}$ that is the relation, which would be produced by $T$ if it's initial state was changed to $\bar{q}$. Note that even though $T$ defines a total relation, $\mathcal{L}(\bar{q})$ might be partial. By $\mathcal{L}(q)$ we denote the relation defined by state in the prefix tree automaton. In particular $\mathcal{L}(q_0)$ for initial state $q_0$ (root of the tree) is equal to the finite part of informant that we are using for learning.

In presence of nondeterminism, the canonical prefix tree automaton could be build in form of a "star". For every sample pair $(\sigma,\gamma)$ from informant we create one path that accepts $\sigma$ and the first transition in the path outputs entire $\gamma$. Then we merge all paths to form one "star" by setting the first state of each path as the root of the tree, which then becomes the initial state.  It's easy to see that every state $q$ in such "star", has a corresponding state $\bar{q}$ in $T$. Only problem with this tree is that there are too many states to merge. If we assume that $T$ locally preserves prefixes, then many of the "star arms" must be merged. Hence we will below introduce a much more efficient form of canonical prefix tree that takes advantage of this property.

For the purpose of our algorithm we need the notion of Brzozowski's derivative but we extend it to regular relations. Given some pair of strings $(\sigma,\gamma)$ and some formal relation $L \subset \Sigma^* \times \Gamma^*$, we can take derivative $(\sigma,\gamma)^{-1}L$ defined as set of all strings in $L$ that begin with $(\sigma,\gamma)$, or more formally $\{(\sigma',\gamma') \in \Sigma^* \times \Gamma^* : (\sigma\sigma',\gamma\gamma') \in L \}$. We also need the $lcp$ function, which given some set of strings, returns their longest common prefix. Functions $\pi_\Sigma(L)$ and $\pi_\Gamma(L)$ are respectively input and output projections of formal relation $L$, which is formally defined as $\pi_\Sigma(L)=\{\sigma\in\Sigma^* : \exists_{\gamma\in\Gamma^*} (\sigma,\gamma)\in L \}$ (analogically for $\pi_\Gamma$).

The prefix tree $P$ is built recursively, starting from the root state. Before we begin the recursion we initialize $q_{i:o}=q_{\epsilon:\epsilon}$ as root of the tree. For any state $q_{i:o}$ of the prefix tree, we define $\mathcal{S}(q_{i:o})$ as the set of all strings in (some finite part of) informant, whose input starts with $i$ and output with $o$. Initially we define $\mathcal{S}(q_{\epsilon:\epsilon})$ as a set containing all the strings in the initial segment of informant that was presented to us. Now we begin the recursion. We check if $(\epsilon,\epsilon)$ belongs to $\mathcal{S}(q_{i:o})$. If it does, we mark $q_{i:o}$ as accepting.  Next for every $\sigma\in\Sigma$, we check if there exists $(\sigma,a)$ in $\mathcal{S}(q_{i:o})$ where $a$ is any string $\Gamma^*$. If it does exist, then we create transition $(q_{i:o},\sigma,q_{i\sigma:oa},a)$ to some new state $q_{i\sigma:oa}$. Then for every symbol $\gamma$ from $\Gamma$, we take the derivative $D = (\sigma,\gamma)^{-1}\mathcal{S}(q_{i:o})$ and compute (if $D$ is not empty) longest common prefix of all possible outputs $lcp(\pi_\Gamma(D))=p$. Check if $oa$ is a prefix of $o\gamma p$ and if it's not, then we create transition $(q_{i:o},\sigma,q_{i\sigma:o\gamma p},\gamma p)\in\delta$ to some new state $q_{i\sigma:o\gamma p}$.  Note that $(\sigma,\gamma p)^{-1}\mathcal{S}(q_{i:o})=\mathcal{S}(q_{i\sigma:o\gamma p})$ and similarly $(\sigma,a)^{-1}\mathcal{S}(q_{i:o})=\mathcal{S}(q_{i\sigma:oa })$. By this point  $q_{i:o}$ may become a leaf (when $D$ was always empty and $q_{i\sigma:oa}$ wasn't created), branch deterministically or nondeterministically. We need to perform the recursion for every outgoing branch, where $q_{i:o}$ becomes each of the newly created state. By the end of running this procedure we have $\mathcal{S}(q_{i:o})=\mathcal{L}(q_{i:o})$ for every state $q_{i:o}$ in the tree. The recursion is well founded, because Brzozowski's derivative gradually shortens strings in $\mathcal{S}(q_{i:o})$ at each recursion level and eventually $D$ will become empty, resulting in $q_{i:o}$ being a leaf state. The procedure also guarantees that if two transitions come out of the same state over the same input symbol, then either A) exactly one of them leads to accepting state or B) neither of them does and common prefix of the two outputs is the empty string.


We can show that for every state $q_2$ in the prefix tree transducer, its incoming edge $e=(q_1,\sigma,q_2,\gamma)$ exactly corresponds to the same edge $\bar{e}=(\bar{q_1},\sigma,\bar{q_2},\gamma)$ in $T$, as soon as all the outgoing edges of $q_2$ have been discovered from the informant. Note that, if some outgoing edges of $q_2$ were missing (not yet known), then the prefix tree transducer might go "too far" in onward form. More precisely, the longest common prefix of all outputs $\Gamma^*$ of all outgoing transitions of $\bar{q_2}$ must be equal to $\epsilon$, but if some of the outgoing transitions of $q_2$ were missing, then their longest common prefix might be a non-empty string $\gamma_2\ne\epsilon$, and it would then be pushed onward to $e=(q_1,\sigma,q_2,\gamma\gamma_2)$. 

We can use the result above to show that for any state $q_{i:o}$ all the outgoing edges of $q_{i:o}$ will be discovered as soon as we read all of the strings $i\Sigma^{\le m}$ where $m$ is the size of target transducer $T$. Assuming that $T$ is trim, every outgoing edge of $q_{i:o}$ will eventually lead to some accepting state. The length of this accepting path can be at most $m$, becasue if it was longer, then by pigeon-hole principle some state would need to repeat and we could find a shorter path without the repetition. Hence if we read all $\Sigma^{\le 2m}$ strings, then we can be sure that all states $q_{i:o}$ with $i<m$ have all of their outgoing edges discovered and their outputs are exactly the same as those of the corresponding edges in $T$.

\section{Inference algorithm}

Inference algorithm is similar to RPNI and OSTIA. We attempt to merge states and look for arising ambiguous paths. Every two ambiguous paths must be unified, until all ambiguity is eliminated. The unification relies on pushing-back outputs whenever necessary. Paths that return different outputs (and thus, break assumption of functional transducer) cannot be unified.  Similarly push-backs that are not transduction-preserving (i.e. a push-back that changes regular relation recognized by transducer) will fail. Merging is also rejected when one ambiguous path contains one of the states that we are trying to merge, but the other one doesn't. If none of the above scenarios occur, and all ambiguous paths are unified, then merge is accepted and inference progresses. The order in which merges are attempted is very important and must be breath-first, or otherwise learning in the limit won't be guaranteed. Below we provide more details.

First we fix an order among states of $P$, such that $q_{i_1:o_1}<q_{i_2:o_2}$ whenever $i_1<_{lex-len}i_2$, where $lex-len$ stands for length-lexicographic order, such that shorter strings are lesser than the longer ones. There are two loops in our algorithm. The outer loop iterates all states $q_{i_2:o_2}$ of $P$ and the inner loop iterates only those states $q_{i_1:o_1}$ that already came earlier that is $q_{i_1:o_1}<q_{i_2:o_2}$. Both loops iterate in increasing $<_{lex-len}$ order. (It's worth pointing out that using Blue-Fringe here would break learning in the limit.) For every pair, of states we attempt to perform their merge. To do this we need to detect ambiguous paths. If merging succeeds, the end result is the deletion of $q_{i_2:o_2}$. The $q_{i_1:o_1}$ retains both states' transitions. 

Checking whether automaton is ambiguous can be done in quadratic time by the squaring procedure\cite{Marie-Pierre}. In order to find the exact paths that are ambiguous, the procedure can be extended to work like a graph search. Squaring of automaton is nothing more than taking its cross product with itself. If $Q$ are states of transducer $P$, then $Q\times Q$ are states of squared automaton $P\times P$. If transducer $P$ has two transitions $(q_1,\sigma,q_2,\gamma_2)$ and $(q_1',\sigma,q_2',\gamma_2')$, then we put a transition $((q_1,q_1'),\sigma,(q_2,q_2'))$ in the squared automaton $P\times P$ (Notice that we lose track of outputs. They are not needed for our purposes). If we  assume that $P$ is trim (all states are reachable and no state is a dead-end) and at any point we encounter a pair $(q_1,q_2)$ in $P\times P$ such that both $q_1$ and $q_2$ transition over the same $\sigma$  to either the same state $q_3$ (formally, there is a transition $((q_1,q_2),\sigma,(q_3,q_3))$ in $P\times P$) or two different accepting states $q_3$ and $q_4$, then we can conclude that $P$ is ambiguous. Hence, finding an ambiguous path reduces to implementing a path-finding algorithm that searches the graph of $P\times P$ for a pair of states $(q_1,q_2)$. 

For additional optimisation, the search can be done incrementally. First we collect all reachable states of $P \times P$. Then as we merge $q_{i_1:o_1}$ with $q_{i_2:o_2}$, we don't "physically merge" them. Instead we scan the set of already reached pairs, and whenever we see $(q_{i_1:o_1},q_2)$ we add $(q_2,q_{i_2:o_2})$ to the set. Similarly if we see $(q_1,q_{i_2:o_2})$ we add $(q_{i_1:o_1},q_1)$. We can also treat the pairs as unordered and this way we don't need to check  $(q_{i_1:o_1},q_1)$ and  $(q_1,q_{i_1:o_1})$ twice. Once we added all those pairs to the set of reachable pairs, we rerun the path-finding procedure and try to discover more reachable states in $P \times P$.

Suppose that the above procedure detected two ambiguous accepting paths 
\begin{equation*}
\begin{split}
(q_0,\sigma_1,q_1,\gamma_1),(q_1,\sigma_2,q_2,\gamma_2),
...(q_{n-1},\sigma_n,q_n,\gamma_n) \\
(q_0,\sigma_1,q_1',\gamma_1'), (q_1',\sigma_2,q_2',\gamma_2'),... (q_{n-1}',\sigma_n,q_n',\gamma_n')
\end{split}
\end{equation*}
that both start in initial state $q_0$. All ambiguous paths must start in $q_0$, because the ambiguity detection procedure allows for "jumps" from $q_{i_1:o_1}$ to $q_{i_2:o_2}$ and vice versa. 
Now in order to ensure that the automaton is unambiguous, we need to merge all the states: $q_1$ with $q_1'$, $q_2$ with $q_2'$, $q_3$ with $q_3'$ and so on. Whenever we encounter $q_{i_1:o_1}=q_k$ or $q_{i_2:o_2}=q_k$ for some $k\le n$ we have to make sure that $q_{i_2:o_2}=q_k'$ or $q_{i_1:o_1}=q_k'$. If that is not the case, then merging must fail immediately.

If all pairs of $q_k$ and $q_k'$ satisfy the above constraint, then we can start unification procedure. The prefix tree automaton $P$ was built in such a way that all transitions are in onward form, but now we might need to push back some of the outputs. We need to ensure $\gamma_1=\gamma_1'$, $\gamma_2=\gamma_2'$... $\gamma_n=\gamma_n'$. Only push back operations are allowed that is, we can "cut off" the suffix from $\gamma_k$ and  prepend it to $\gamma_{k+1}$, but we are not allowed to "cut off" prefix of $\gamma_{k+1}$ and append it to $\gamma_k$. Similarly for $\gamma_k'$ and $\gamma_{k+1}' $. Attention must be paid as some push-backs might alter the regular relation recognized by transducer. For simplicity, we ensure that all push-backs are transduction-preserving  by only allowing them, when state has only one incoming transition and is not accepting. That is, a suffix of $\gamma_k$ can become a prefix of $\gamma_{k+1}$ under the condition that state $q_k$ has one single incoming transition $(q_{i-1},\sigma_k,q_k,\gamma_k)$ and $q_k$ is not accepting. Therefore, sometimes path unification may fail and some states may not be merged.
State merging might also fail if it breaks functionality of transducer that is, if it there are two accepting paths but their outputs are different $\gamma_1\gamma_2...\gamma_n\ne\gamma_1'\gamma_2'...\gamma_n'$.

\section{Proof of learning in the limit}

Let $P$ be the prefix tree transducer. We assume that the target transducer $T$ is functional, unambiguous, trim, total (except for $\epsilon$) and its transitions satisfy the following properties:

Suppose that $T$ has $m$ states and that we saw all of the inputs $\Sigma^{\le 2m}$ from informant. Hence for every $q_{i_1:o_1}$ in $P$ such that $\vert i_1\vert < m$,  if there exists an edge $(\bar{q}_{i_1:o_1},\sigma,q_2,\gamma)$ in $T$, then there must exist $(q_{i_1:o_1},\sigma,q_{i_1\sigma:o_1\gamma},\gamma)$ in $P$ such that $\bar{q}_{i_1\sigma:o_1\gamma}=q_2$. We will refer to every transition  $(q_{i_1:o_1},\sigma,q_{i_1\sigma:o_1\gamma},\gamma)$ in $P$ using the unique label $e_{i_1\sigma:o_1\gamma}$. The above observation can be restated as: for every $e_{i_1\sigma:o_1\gamma}$ such that $\vert i_1\sigma\vert \le m$ there exists $\bar{e}_{i_1\sigma:o_1\gamma}$ in $T$ whose output is exactly $\gamma$ and target state of $e_{i_1\sigma:o_1\gamma}$ corresponds to target state of $\bar{e}_{i_1\sigma:o_1\gamma}$.

When we merge two states $q_{i_1:o_1}$ and $q_{i_2:o_2}$ in $P$ such that $\vert i_1 \vert < m$ and $\vert i_2 \vert < m$, there might arise many ambiguous paths and many transitions need to be unified. Suppose that $\bar{q}_{i_1:o_1}=\bar{q}_{i_2:o_2}$ and we unify $e_{i_3:o_3}$ with $e_{i_4:o_4}$. There are four cases. 
\begin{enumerate}
	\item Suppose $\vert i_3 \vert \le m $ and  $\vert i_4 \vert \le m $ then their outputs must be exactly equal and no push-back is necessary.
	\item Suppose $\vert i_3 \vert \le m $ and  $\vert i_4 \vert > m $ then the output of $e_{i_4:o_4}$ might require to be pushed-back and as a result it will become equal to output of $e_{i_3:o_3}$ and $\bar{e}_{i_3:o_3}$. Hence $e_{i_4:o_4}$ will never need to be pushed-back again.
	\item  $\vert i_3 \vert > m $ and  $\vert i_4 \vert \le  m $ is same as above
	\item Suppose $\vert i_3 \vert > m $ and  $\vert i_4 \vert > m $ then push-backs might be necessary on either side and they are not guaranteed to result in the exact same output as in $T$. However, it is guaranteed that $e_{i_3:o_3}$ is the only transition incoming to its target state and similarly is $e_{i_4:o_4}$ (because inference algorithm will never attempt to merge targets of two such edges). Hence after we unify target of $e_{i_3:o_3}$ with target of $e_{i_4:o_4}$, the resulting state will also have only one incoming transition. Therefore all future push-backs on this transition will be transduction-preserving and eventually the correct transition output will be inferred. 
\end{enumerate}

Because transducer is total, this proves that if we are able to correctly identify and merge all those states $q_{i_1:o_1}$ and $q_{i_2:o_2}$ with $\vert i_1 \vert < m$ and $\vert i_2 \vert < m$, then the transducer will be fully inferred before we have the chance to merge any states further than $m$.

In order to prove that only and all the correct merges will be performed, we first prove that $q_{i_1:o_1}$ will be merged with $q_{i_2:o_2}$ only if $\mathcal{L}(\bar{q}_{i_1:o_1})=\mathcal{L}(\bar{q}_{i_2:o_2})$ and that no correct merge will be mistakenly missed. There are the following cases.

\begin{enumerate}
	\item Suppose that $\mathcal{L}(\bar{q}_{i_1:o_1}) \cup \mathcal{L}(\bar{q}_{i_2:o_2})$ is not a function.
	Then there exists some string $i\in\Sigma^*$ such that $(i,o) \in \mathcal{L}(\bar{q}_{i_1:o_1})$ and $(i,o') \in \mathcal{L}(\bar{q}_{i_2:o_2})$ and that $o\ne o'$. We need to wait until informant shows us the two examples $(i_1 i, o_1 o)$  and $(i_2 i, o_2 o')$ and then merging $q_{i_1:o_1}$ with $q_{i_2:o_2}$ will become impossible, because the two paths will be ambiguous and have outputs impossible to unify. 
	
	\item Suppose that $\mathcal{L}(\bar{q}_{i_1:o_1}) \subset \mathcal{L}(\bar{q}_{i_2:o_2})$ holds and $\mathcal{L}(\bar{q}_{i_2:o_2}) \subset \mathcal{L}(\bar{q}_{i_1:o_1})$ doesn't. Then there will exist some $(i,o')\in \mathcal{L}(\bar{q}_{i_2:o_2})$ such that 
	$i$ is not present in $\mathcal{L}(\bar{q}_{i_1:o_1})$.	
	Because transducer is total, there must exist (in the limit) some state $q_{i_1:o_3}$ such that $(i,o)\in \mathcal{L}(q_{i_1:o_3})$. If we attempt to merge   $q_{i_1:o_1}$ with $q_{i_2:o_2}$ we will detect two ambiguous paths $(i_1 i, o_3 o')$ and $(i_1 i , o_3 o)$. Those paths will be respectively 

	\begin{equation*}
	\begin{split}
	(q_0,\sigma_1,q_1,\gamma_1),...
	(q_{k-1},\sigma_k,q_{i_2:o_2},\gamma_k),...
	(q_{n-1},\sigma_n,q_n,\gamma_n) \\
	(q_0,\sigma_1,q_1',\gamma_1'),... (q_{k-1}',\sigma_k,q_{i_1:o_3},\gamma_k'),... (q_{n-1}',\sigma_n,q_n',\gamma_n')
	\end{split}
	\end{equation*}

	and the inputs are $i_1=\sigma_1\sigma_2...\sigma_k$ and $i=\sigma_{k+1}...\sigma_n$. We can see that $q_k=q_{i_2:o_2}$ but $q_k'$ is neither $q_{i_1:o_1}$ nor $q_{i_2:o_2}$, hence merging will be rejected.
	
	\item Suppose that $\mathcal{L}(\bar{q}_{i_2:o_2}) \subset \mathcal{L}(\bar{q}_{i_1:o_1})$ holds and  $\mathcal{L}(\bar{q}_{i_1:o_1}) \subset \mathcal{L}(\bar{q}_{i_2:o_2})$ doesn't, then do analogically as above.
\end{enumerate}

Hence we proved that, in the limit, merges will be done only if $\mathcal{L}(\bar{q}_{i_1:o_1})=\mathcal{L}(\bar{q}_{i_2:o_2})$. Now we need to prove that no correct merges will be missed. A merge can be rejected for 3 reasons:

\begin{enumerate}
	\item Path cannot be unified, because some push-back is not transduction-preserving due to multiple incoming transitions. We already proved above that this is not an issue.
	\item Path cannot be unified, because some push-back is not transduction-preserving due to some state being accepting too early. But if this happens, then the merge cannot possibly be valid, because we started in onward form and all the push-back we've done so far were valid (by induction).
	\item Path cannot be unified, because the outputs are different and break functionality. If this happens, then the merge cannot be correct.
	\item Path cannot be unified because for some $k$ the state $q_k$ is either  $q_{i_1:o_1}$ or $q_{i_2:o_2}$ but the state  $q_k'$ is neither of those. Let's prove that this can never be a correct merge.
	
	We define configuration $K$ as any subset of $Q \rightarrow \Gamma^*$. By $K_i$ we denote the configuration reached after reading input $i\in\Sigma^*$. Formally $K_\epsilon$ is the singleton set $\{(q_0,\epsilon)\}$ and recursive definition of $K_{i \sigma}$ is $\{(q_2,o\gamma) \in  Q \rightarrow \Gamma^* : \exists_{(q_1,o)\in K_i} (q_1,\sigma,q_2,\gamma)\in\delta \}$. 
	
	If any two states $q$ and $q'$ in $P$ such that $\bar{q}=\bar{q}'$,  belong to $K_i$ for some $i\in\Sigma^*$,  then $q=q'$, because otherwise there would exist two different paths over $i$ in $T$ that both lead to $\bar{q}$ and $T$ would not be functional (and transitions of $P$ guarantee us that if there are two different paths over $i$ then they have distinct outputs that are not prefixes of one another). This guarantees us that as we attempt to merge $q_{i_1:o_1}$ with $q_{i_2:o_2}$, then for any $K_i$, the $q_{i_1:o_1}$ is in $K_i$ if and only if $q_{i_2:o_2}$ is in $K_i$ and those are the only states in $K_i$ that correspond to $\bar{q}_{i_1:o_1}$. 
	
\end{enumerate}

To finish the proof we need to conclude that $\mathcal{L}(\bar{q}_{i_1:o_1})=\mathcal{L}(\bar{q}_{i_2:o_2})$ implies $\bar{q}_{i_1:o_1})=\bar{q}_{i_2:o_2}$. This holds true because, if it didn't we could find another transducer equivalent to $T$, but smaller, by deleting $\bar{q}_{i_1:o_1}$ and redirecting all the transition incoming to $\bar{q}_{i_1:o_1}$ as incoming to $\bar{q}_{i_2:o_2}$ instead (this would not violate any restrictions we imposed earlier on the transitions of $T$).

\section{Conclusions}
This concludes the description of onward functional transducer inference algorithm. There is one interesting thing we would like to point out. One could say that RPNI is a special case of OSTIA, where the output is always the empty string. In particular it should be observed that every finite state automaton is a special case of finite state transducer that either rejects (prints null $\emptyset$ output) or accepts (prints empty $\epsilon$ output). The algorithm described in this paper is a "superalgorithm" that can behave like OSTIA when the target transducer is deterministic (and outputs preserve their prefixes). 
Moreover, note that there is no need to introduce subsequential transducers, because the state output can be simulated with non-determinism. The only exception being the state output of initial state. This limitation is not a problem, because the output generated by initial state (output associated with empty input string), can be learned independently. More precisely, as soon as informant shows us the output associated with $\epsilon$, we save it somewhere aside and then learn the transducer as usual, by pretending that $(\epsilon,\epsilon)$ belongs to the regular relation.

\ifCLASSOPTIONcaptionsoff
  \newpage
\fi



%




\bibliographystyle{BibTeXtran}   
\bibliography{BibTeXrefs}       

%







\end{document}